\documentclass[english,twocolumn,showkeys]{article}
\usepackage[
  a4paper,
  textwidth=179.5mm,
  columnsep=10pt
]{geometry}
\usepackage{amsmath,graphicx}
\usepackage{textcomp}
\usepackage{amssymb}
\usepackage{siunitx}
\usepackage{pdfpages}
\usepackage{changes}
\usepackage{float}
\usepackage{braket}
\usepackage{dblfloatfix}   
\usepackage{balance}
\usepackage[version=4]{mhchem}
\usepackage{pifont}
\usepackage{cuted}
\usepackage{bm}
\usepackage[sorting=none,style=nature,doi=true,isbn=false,url=false,eprint=false,date=year,pluralothers=false, maxnames=100]{biblatex}
\usepackage[hidelinks,colorlinks=true,linkcolor=blue,citecolor=blue]{hyperref}
\usepackage{enumitem}

\newcommand{\hema}{$\alpha$-\ce{Fe2O3}}
\newcommand{\TM}{$T_\mathrm{M}$}

\usepackage{authblk}
\title{Second-Harmonic Imaging of Magnetic Domains in Thin Film Hematite}

\newcommand{\equalcontrib}{\textsuperscript{*}}
\newcommand{\mail}{\textsuperscript{$\dagger$}}
\author[1,2]{Holger Mirkes\equalcontrib}
\author[1,2]{Johannes Schmuck\equalcontrib}
\author[3,4]{Katharina Müller}
\author[5]{János Papp}
\author[5]{Paul Seifert}
\author[3,4]{Matthias Althammer}
\author[3,4,2]{Hans Huebl}
\author[3]{Stephan Geprägs}
\author[1,2]{Alexander W. Holleitner}
\author[1,2]{Christoph Kastl\mail}

\affil[1]{Walter Schottky Institute and Physics Department, Technical University of Munich, Am Coulombwall 4a, 85748 Garching, Germany}
\affil[2]{Munich Center of Quantum Science and Technology (MCQST), Schellingstraße 4, 80799 Munich, Germany}
\affil[3]{Walther-Meißner-Institut, Bayerische Akademie der Wissenschaften, Walther-Meißner-Straße 8, 85748 Garching, Germany}
\affil[4]{Physics Department, TUM School of Natural Sciences, Technical University of Munich, 85748 Garching, Germany}
\affil[5]{Institute of Physics and Center for Integrated Sensor Systems (SENS), University of the Bundeswehr Munich, Werner-Heisenberg-Weg 39, 85577 Neubiberg, Germany}

\date{\equalcontrib These authors contributed equally to this work.\\[0.5ex]\mail Email: christoph.kastl@wsi.tum.de}

\begin{document}

\maketitle
\begin{abstract}

Hematite is an antiferromagnetic oxide and candidate altermagnetic insulator whose N\'eel order reorients from an easy-axis to an easy-plane phase at the Morin transition. Interpreting altermagnetic transport and symmetry-sensitive optical responses requires knowledge of the N\'eel-vector orientation relative to the crystal axes and of the domain structure within the probed device. Here, we show that polarization-resolved second-harmonic generation (SHG) microscopy resolves magnetic symmetry and domains in epitaxial (0001)-oriented hematite films. Across the Morin temperature, the SHG polarization anisotropy evolves from an approximately sixfold pattern consistent with the easy-axis orientation to pronounced twofold patterns consistent with the easy-plane orientation. A symmetry analysis based on magnetic-dipole and electric-quadrupole contributions reproduces this evolution. Importantly, the interference between SHG amplitudes that are odd and even under reversal of the magnetic order renders opposite N\'eel-vector orientations optically distinguishable. Consistently, opposite directions of an applied in-plane magnetic field produce distinct SHG responses in our experiments. Using this magnetic contrast, we image micrometer-scale domains, their reorganization across the Morin transition, and their reconfiguration under magnetic and thermal cycling, including a remanent change after cycling through the spin-flop transition. These results establish SHG microscopy as a local probe of magnetic symmetry, N\'eel-vector orientation, and domain evolution in hematite films.

\end{abstract}

\section*{Keywords}
Second-Harmonic Generation, Hematite, Antiferromagnetism, Altermagnetism, Magnetic Domains, Non-linear Optics

\section{Introduction}

Compensated magnets promise improved spintronic performance compared to ferromagnets \cite{althammer_all-electrical_2021, han2023coherent}: they are robust against external magnetic field perturbations, and they allow for fast magnetization dynamics governed by strong inter-sublattice exchange rather than weak anisotropy fields \cite{baltz2024emerging}. The identification of altermagnetic order has recently enriched the landscape of compensated magnets \cite{vsmejkal2022emerging, smejkal2022prx}. In contrast to conventional antiferromagnets, whose electronic bands remain spin-degenerate despite broken time-reversal symmetry, apart from small spin-orbit-induced splittings, altermagnets exhibit a strong momentum-dependent spin and magnon splitting that alternates between specific crystallographic directions \cite{jungwirth2025altermagnetism, altermagnetic_spintronics}.

\begin{figure*}[htbp]
\centering
\includegraphics[width=0.95\textwidth]{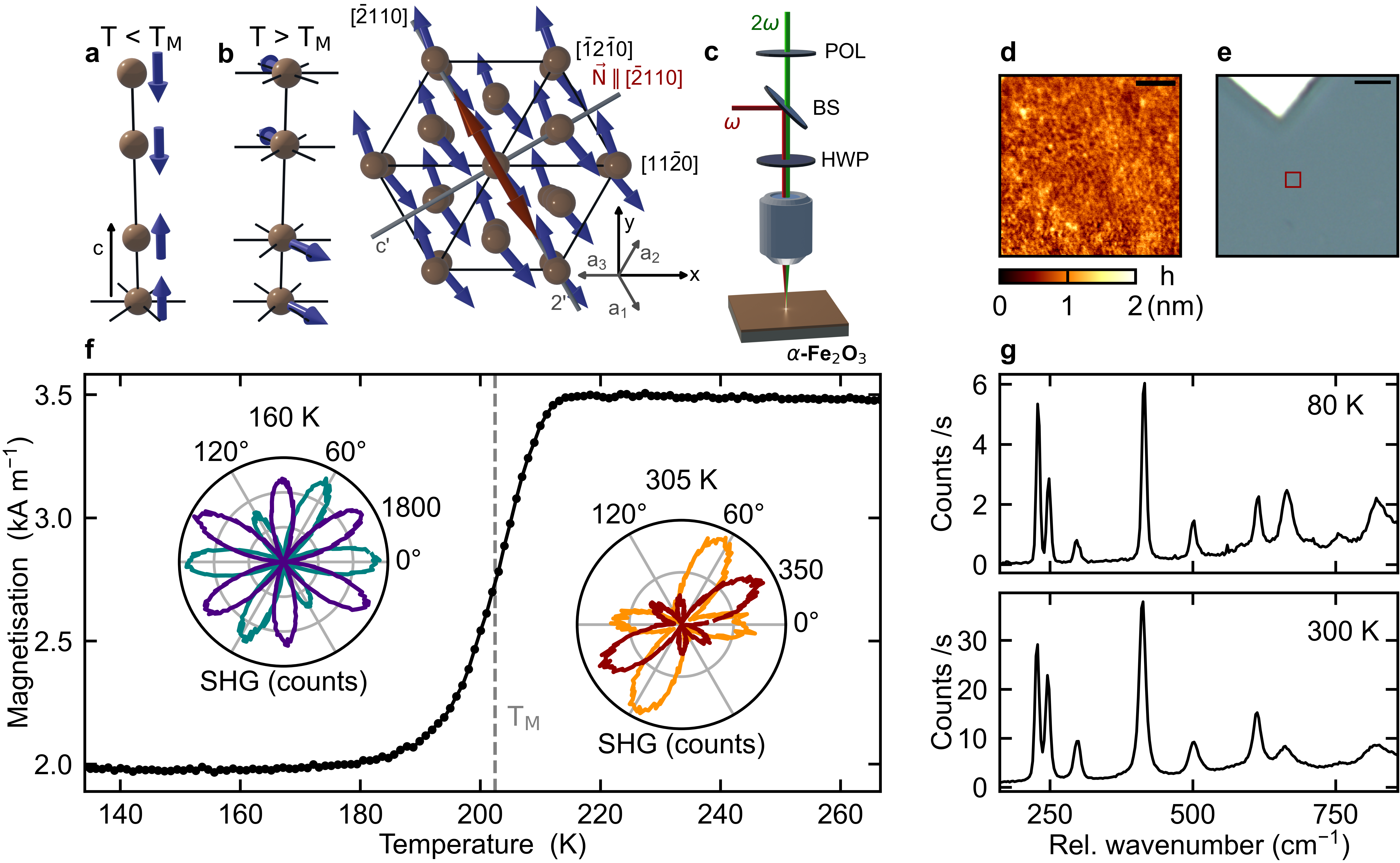}
\caption{\textbf{Structural and magnetic properties of hematite (\hema) thin films.} \textbf{a} Below the Morin temperature (\TM{}), spins are aligned in an easy-axis configuration along the [0001] axis. \textbf{b} For $T_\mathrm{M} < T < T_\mathrm{N}$, the spins reorient into an easy-plane configuration with a small canting caused by DMI. The resulting weak ferromagnetic moment points along the mirror axes $c'$. The N\'eel vector is along the perpendicular direction (red) and parallel to the two-fold axis $2'$. The shown coordinate system defines the relative orientation between the lab reference system (xy-coordinates) and the in-plane crystal orientation (hexagonal lattice vectors $\mathbf{a}_1$, $\mathbf{a}_2$, $\mathbf{a}_3$). \textbf{c} Sketch of the SHG setup with a rotatable half-wave plate (HWP), dichroic beamsplitter (BS), and analyzing polarizer in the detection path (POL). \textbf{d} Topography of an epitaxial \qty{89}{nm} thin \hema{} film grown on (0001)-oriented sapphire (scale bar: \qty{200}{nm}). \textbf{e} Optical microscope image of the thin film sample with metallic markers shown in white contrast. The red box indicates the area in which SHG measurements are recorded (scale bar: \qty{50}{\micro\meter}). \textbf{f} Magnetometry reveals a Morin transition at \TM{} $\approx \qty{203}{K}$. Insets show the polarization-resolved SHG intensity at \qty{160}{K} and \qty{305}{K} in co-polarization (purple and red) and cross-polarization (petrol and orange). The origin of the polar plots corresponds to zero counts. \textbf{g} Raman spectra recorded at \qty{80}{K} and \qty{300}{K}.}\label{fig1}
\end{figure*}

Hematite (\hema{}) has emerged as an altermagnetic candidate with a predicted and experimentally confirmed chiral magnon splitting \cite{hoyer2025altermagnetic, sun2025observation} and an experimentally observed crystal direction dependent magnetotransport in doped crystals. Undoped hematite is an electrical insulator and crystallizes in a corundum structure with space group $R\overline{3}c$. Below the Néel temperature $T_\mathrm{N} \approx \qty{950}{K}$ \cite{bialek2022antiferromagnetic}, it orders magnetically. Upon cooling, hematite undergoes a spin-reorientation transition from the high-temperature easy-plane configuration to the low-temperature easy-axis configuration. In bulk crystals, this so-called Morin transition occurs at $T_\mathrm{M} = \qty{263}{K}$ \cite{morin_magnetic_1950}. For $T < T_\mathrm{M}$, the magnetic moments align collinearly along the crystallographic $c$-axis (Fig.~\ref{fig1}\textbf{a}), while for $T_\mathrm{M} < T < T_\mathrm{N}$, the Néel vector lies in the basal plane with three distinct high-symmetry orientations (Fig.~\ref{fig1}\textbf{b}). In the latter phase, the Dzyaloshinskii--Moriya interaction (DMI) induces a weak canting and a small net magnetization \cite{thoma2021}.
Several recent works have demonstrated the scalable growth of high-quality epitaxial \hema{} thin films \cite{geprags2020spin,scheufele_impact_2023,Liu2025anisotropic, toda2025substrate} with promising spintronic functionality \cite{wimmer_observation_2020,fischer2020large, gueckelhorn2023, fritjofson2025coherent}. A challenge remains the coexistence of different easy-plane domain orientations \cite{fabian2011experimental}. In this case, spin and electronic transport as well as magnetometry \cite{hamdi2023spinwave, lebrun2020long, martin2013evidence} often probe a superposition of domains, complicating the interpretation of observables that depend on the specific magnetic symmetry or orientation of the Néel vector. Magnetic domains \cite{jani_antiferromagnetic_2021} as well as current-induced switching of the Néel vector \cite{cogulu_direct_2021} were investigated using X-ray magnetic linear dichroism in \hema{} films covered with a thin Pt layer. Recent work further connected the symmetry of the Néel vector to the anomalous Hall effect, consistent with an expected altermagnetic spin splitting \cite{galindez2025revealing}. In these studies, the reported lateral size of domains in the easy-plane phase is on the order of micrometers.

In this respect, second harmonic generation (SHG)\cite{fiebig_second-harmonic_2005, ni_magnetic_2024, ho_imaging_2025} can provide an alternative to the synchrotron-based magnetic imaging employed so far\cite{jani_antiferromagnetic_2021,cogulu_direct_2021, galindez2025revealing}.

The magnetic sensitivity of SHG is governed by nonlinear susceptibility tensors whose components linked to both the structural and magnetic point group \cite{fiebig_second-harmonic_2005}. 
Polarization-resolved SHG provides access not only to magnetic symmetries, but also to domain contrast with diffraction limited resolution \cite{fiebig1996, fiebig_second_2001, fiebig_second-harmonic_2005, zhao_evidence_2016, xu2023magnetoelectric, ni_magnetic_2024, wang2024electric,zhang2026magnetic,ho_imaging_2025}. This is relevant in antiferromagnets, where domains may be invisible to linear magneto-optical probes and where structurally equivalent domains can become distinguishable in nonlinear response once (alter)magnetic point-group symmetries are considered \cite{nemec_antiferromagnetic_2018}. Beyond real-space imaging, SHG can also be used spectroscopically. For example, tuning the fundamental photon energy through electronic resonances can selectively enhance nonlinear tensor components associated with band-to-band transitions or localized orbital excitations providing sensitivity to magnetic-order-induced changes in electronic structure, local crystal fields, or subtle reductions of point-group symmetry \cite{fiebig_second-harmonic_2005}.

Here, we investigate the evolution of magnetic domains in epitaxial \hema{} thin films across the Morin transition by polarization-resolved SHG (Fig. \ref{fig1}\textbf{c}). Above $T_\mathrm{M}$, the SHG shows a distinct polarization anisotropy with micrometer-sized lateral domains, which we attribute to symmetry-equivalent in-plane orientations of the Néel vector. Upon cooling, the SHG intensity increases and the domain patterns reorganize at the Morin transition temperature. Moderate in-plane magnetic fields above \TM{} and large out-of-plane magnetic fields below \TM{} modify the SHG patterns, confirming their magnetic origin. Due to the centrosymmetric structure of hematite, SHG originates from otherwise weak higher-order electric and magnetic contributions. A symmetry analysis using magnetic-dipole and electric-quadrupole terms, assumed to be comparable in magnitude, consistently describes the observed polarization dependencies and magnetic contrast. Our results establish polarization-resolved SHG as a local probe of magnetic symmetry, domain structure, and device homogeneity in \hema{}, enabling, for example, transport studies on microscopic devices with known or homogeneous magnetic properties.

\section{Results and Discussion}

\begin{figure}[htbp]
\centering
\includegraphics[width=0.90\columnwidth]{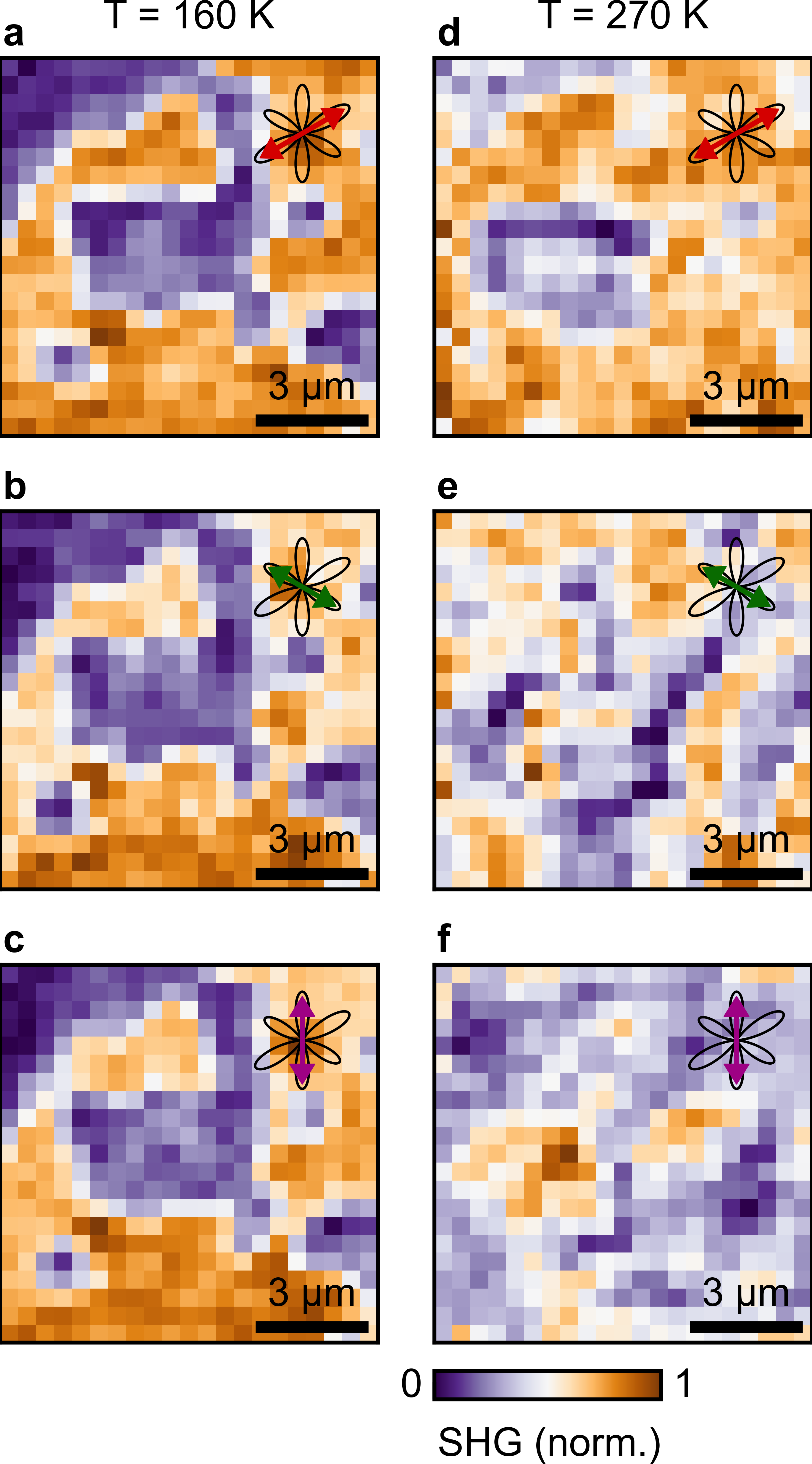}
\caption{\textbf{Spatial SHG mapping of magnetic domains.} \textbf{a}-\textbf{c} SHG images recorded at \qty{160}{K} below \TM. \textbf{d}-\textbf{f} SHG images recorded at \qty{270}{K} above \TM. The measurements are taken with the polarization angle set to \qty{30}{\degree}, \qty{150}{\degree} and \qty{270}{\degree}, as indicated by the insets. All maps are normalized individually.}\label{fig2}
\end{figure}

\begin{figure*}[htb]
\centering
\includegraphics[width=0.95\textwidth]{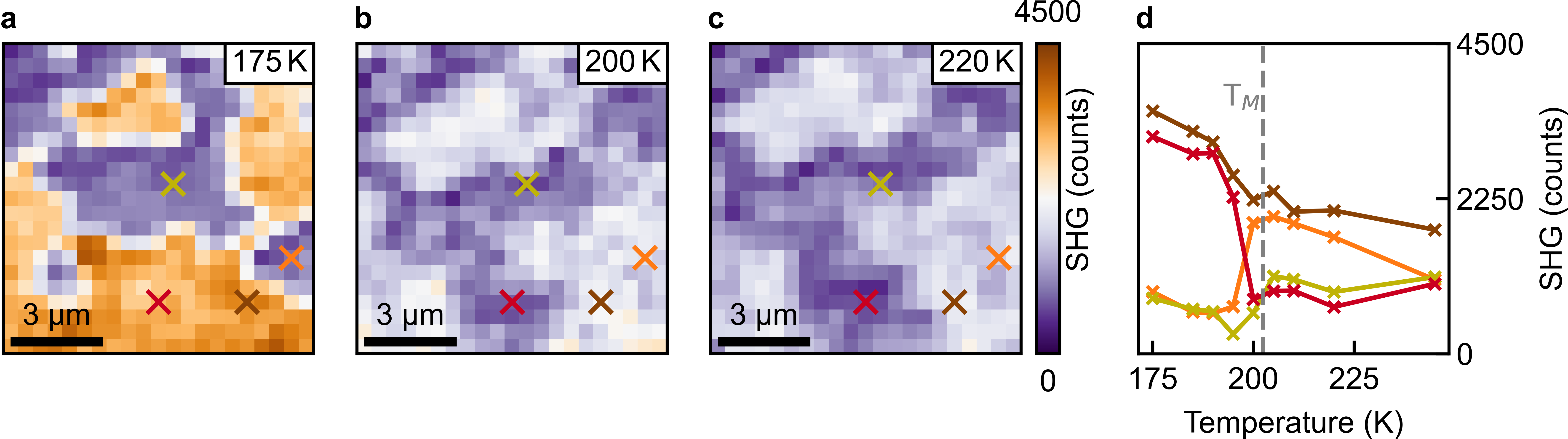}
\caption{\textbf{Confocal images of the SHG intensity.} Images are recorded at \textbf{a} \qty{175}{K}, \textbf{b} \qty{200}{K}, and \textbf{c} \qty{220}{K} with a co-polarized detection for \qty{30}{\degree} and a laser power of $P_\mathrm{laser} = \qty{100}{mW}$.
\textbf{d} Temperature dependence of the SHG intensity at the sample positions indicated in \textbf{a-c}. The SHG intensity changes distinctly across the Morin transition.}\label{fig3}
\end{figure*}

\begin{figure}[htb]
\centering
\includegraphics[width=0.95\columnwidth]{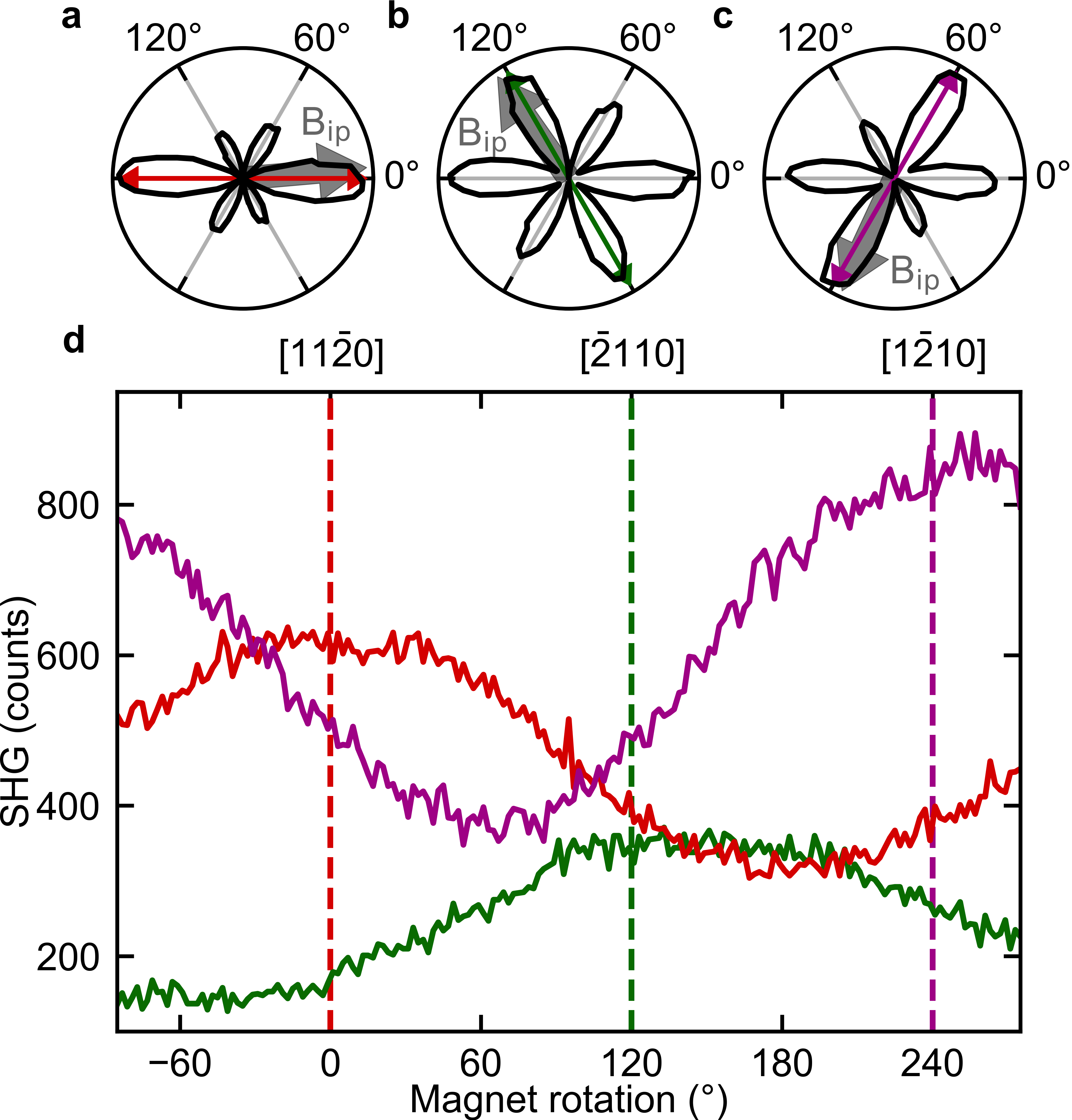}
 \caption{\textbf{Polarization-dependent SHG with rotated in-plane magnetic field.} \textbf{a-c} SHG at a fixed sample position and at room temperature for different directions of the in-plane magnetic field $B_\mathrm{ip}= \qty{550 \pm 50}{mT}$ detected in cross-polarization. \textbf{d} The SHG intensity along the three different polarizations defined by colored arrows in a-c varies continuously as function of the in-plane field angle. Importantly, the SHG intensity differs for opposite field directions.}\label{fig4}
\end{figure}

\begin{figure*}[htbp]
\centering
\includegraphics[width=0.9\textwidth]{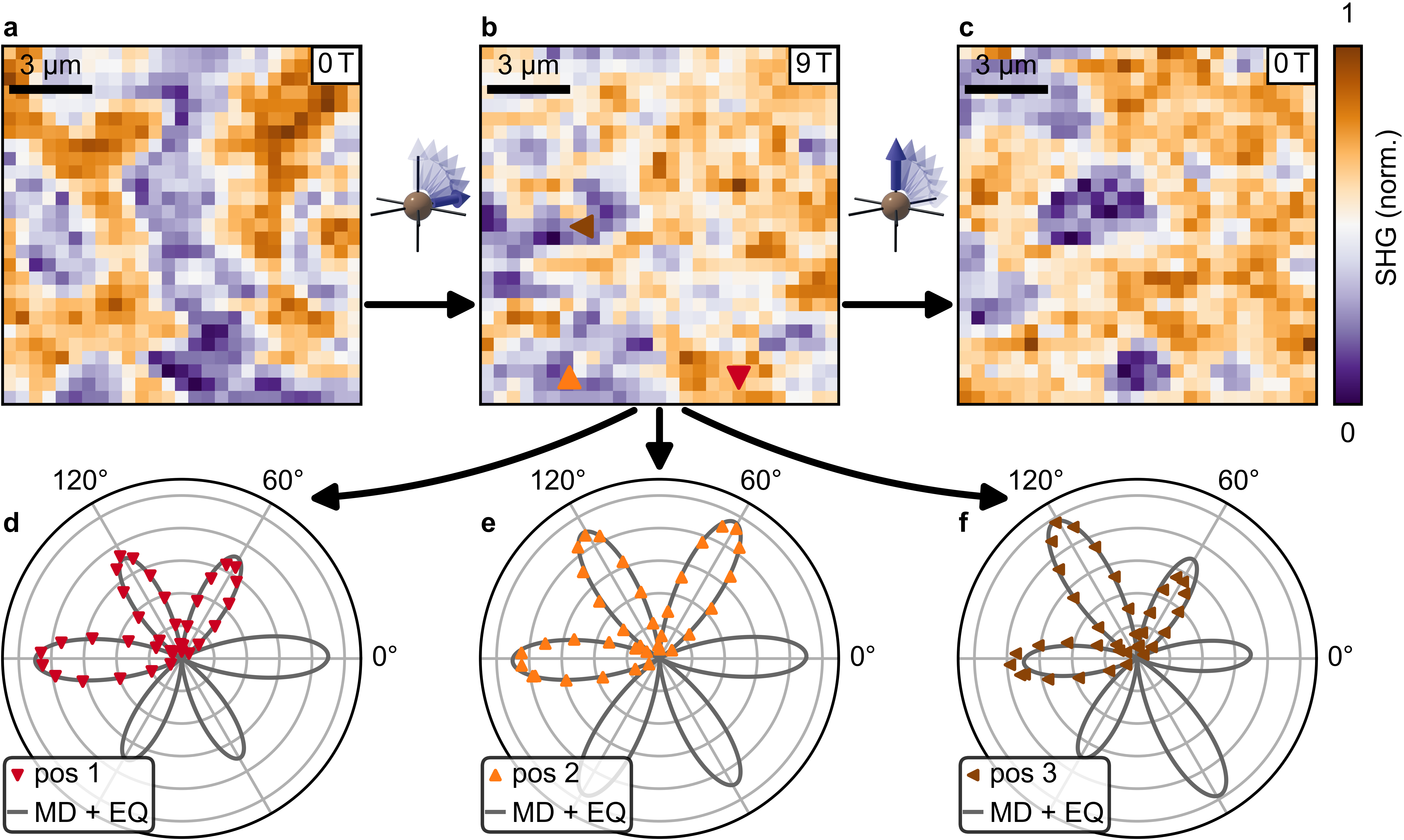}
\caption{\textbf{Spatial SHG mapping of a spin-flop induced by an out-of-plane magnetic field.} Consecutive SHG images recorded at \textbf{a} \qty{0}{T} out-of-plane magnetic field, \textbf{b} \qty{9}{T} out-of-plane magnetic field (above the spin-flop transition), \textbf{c} again \qty{0}{T} out-of-plane magnetic field. \textbf{d-f} Rotational SHG patterns at the positions indicated in \textbf{b}. The data points (triangles) are symmetrized to account for temperature drifts during the measurements. The gray solid line is a fit using a magnetic dipole (MD) and electric quadrupole (EQ) source term. For all measurements $T = \qty{4.7}{K}$, $P_\mathrm{laser} = \qty{70.3}{mW}$, cross-polarized detection.}\label{fig5}
\end{figure*}

The epitaxial \hema{} films investigated in the present study are grown via pulsed laser deposition on (0001)-oriented, single crystalline sapphire (\ce{Al2O3}) substrates (see methods) \cite{scheufele_impact_2023}. Figure~\ref{fig1}\textbf{d} exemplarily shows an atomic force microscopy topography image of a 89\,nm thick \hema{} thin film revealing a RMS roughness of 0.82\,nm demonstrating the smoothness of the \hema{} thin films. 
Figure~\ref{fig1}\textbf{e} depicts an optical microscopy image, where the red box indicates the approximate location of the SHG measurements discussed later in Fig. \ref{fig2} through Fig. \ref{fig5}. The positions of all measurements are referenced to metal markers (visible as white contrast in Fig. \ref{fig1}\textbf{e}) with a precision better than \qty{1}{\micro\meter}. First, we verify by magnetometry that the film exhibits a Morin transition, which is highly dependent on the correct growth conditions \cite{scheufele_impact_2023,Liu2025anisotropic}. The magnetization curve indicates a clear Morin transition at \TM{} = \qty{203}{K} (Fig.~ \ref{fig1}\textbf{f}), which is lower than typical values for bulk crystals \cite{bialek2022antiferromagnetic} but consistent with previous reports on thin epitaxial films \cite{Liu2025anisotropic}. To corroborate the high quality of our films and verify the absence of any structural lattice changes across the observed Morin transition, that may impact the SHG, we show Raman spectra at \qty{80}{K} and \qty{300}{K} (Fig.~\ref{fig1}\textbf{g}). The spectra agree very well with typical Raman modes reported for single crystal bulk hematite (see Supplemental Figure S1 for a detailed analysis)\cite{lopez-sanchez_large_2022, marshall_polarized_2020}.

The crystal structure of hematite is centrosymmetric, which forbids the electric dipole contribution to SHG \cite{pisarev_nonlinear_1997}. We therefore consider a magnetic dipole (MD) and an electric quadrupole (EQ) source term as the leading order contributions in our symmetry analysis (for details see Supplemental Material) \cite{gallego2019automatic,Wu2022,guo_ferrorotational_2023}.
We assume normally incident light propagating along the \(c\)-axis, $E=E_0(\cos\phi,\sin\phi,0)$. For all following measurements and calculations, the lab reference system (referenced with xy-coordinates) and the in-plane crystal orientation (referenced with hexagonal lattice vectors $\mathbf{a}_1$, $\mathbf{a}_2$, $\mathbf{a}_3$) are aligned as sketched in Fig. \ref{fig1}\textbf{b}. At a polarization angle $\phi =\qty{0}{\degree}$, the incident light is polarized along the direction of a hexagonal lattice vector, such as the $[11\overline{2}0]$ direction. In this configuration, we need to distinguish the following two cases. Below \TM, the Néel vector is collinear with the \(c\)-axis, and it does not break the in-plane threefold rotational symmetry. As a result, both the electric (EQ) and magnetic (MD) SHG contributions retain their trigonal symmetry with $\bar{3}m$ and $\bar{3}^\prime m^\prime$, respectively \cite{pisarev_nonlinear_1997}. Above \TM, the Néel vector lies within the (0001)-plane. The DMI induced canting effectively reduces the symmetry of the MD contribution to $2'/m'$, where the small ferromagnet moment arising from the canting is parallel and the N\'eel vector is perpendicular to the mirror axis $c'$ (cf. Fig. \ref{fig1}\textbf{b}) \cite{galindez2025revealing, hoyer2025altermagnetic, ishii2026altermagnetic}. Projecting the symmetry-allowed nonlinear susceptibilities ($\chi^m_{ijk}$ and $\chi^q_{ijkl}$) onto the co-polarized ($I_{XX}$) and cross-polarized detection channels ($I_{YX}$), we find for the expected SHG intensity patterns 
\begin{equation}
    \begin{aligned}
        I^{T<T_M}_{XX}\propto&|\chi^m_{xxx}\sin3\phi+\chi^q_{xxzy}\sin3\phi|^2,\\ 
        I^{T<T_M}_{YX}\propto&|\chi^m_{xxx}\cos3\phi+\chi^q_{xxzy}\cos3\phi|^2,\\ 
        I^{T>T_M}_{XX}\propto&|-2\chi^m_{xyy}\cos^2\phi \sin\phi+\\
        &+\chi^m_{xxx}\sin\phi\cos^2\phi+\chi^m_{xyy}\sin^3\phi+\chi^q_{xxzy}\sin3\phi|^2,\\ I^{T>T_M}_{YX}\propto&|3\chi^m_{xyy}\sin^2\phi\cos\phi+\chi^m_{xxx}\cos^3\phi+\chi^q_{xxzy}\cos3\phi|^2. 
    \end{aligned}
\end{equation}

Hence, in cross-polarized measurements, the lobes of the SHG signal must be aligned with the $[11\overline{2}0]$, $[\overline{1}2\overline{1}0]$, $[\overline{2}110]$ crystal axes (c.f. Fig. \ref{fig1}\textbf{b} and Supplemental Figure S2). Conversely, in co-polarized measurements, the lobes must be oriented in between these crystal axes. In the following, we demonstrate that these source terms consistently describe the rotational dependence of the SHG patterns above and below \TM{} as well as the magnetic contrast. 

The insets in Fig.~\ref{fig1}\textbf{f} depict the measured SHG intensity as a function of linear excitation polarization (excitation wavelength \qty{1064}{nm}, detection wavelength \qty{532}{nm}, see Fig.~\ref{fig1}\textbf{c} and methods for details) both below \TM{} measured at \qty{160}{K} (purple and petrol) and above \TM{} at \qty{305}{K} (red and orange). As expected from the symmetry analysis above, the maxima of the cross-polarized signal (petrol and orange lines) align with the hexagonal lattice vector directions ($n\cdot \qty{60}{\degree}$ in the angular reference frame), and the maxima of the co-polarized signal (purple and red lines) are in between them ($\qty{30}{\degree}+n\cdot \qty{60}{\degree})$. In the easy-plane phase (above \TM), both SHG patterns clearly reflect the reduced symmetry of the magnetic point group with a two-fold symmetric lobe pattern (red and orange line). As discussed below, the shape of this pattern depends on the in-plane magnetization direction, which changes either in space through domain formation or in an external in-plane magnetic field through N\'eel vector alignment. In the easy-axis phase (below \TM), the SHG pattern becomes approximately sixfold symmetric (purple line in Fig. \ref{fig1}\textbf{f}), as would be expected based on the overall trigonal symmetry. We note that we still find a residual distortion of the SHG pattern, which becomes more evident in the cross-polarized signal (petrol line in Fig. \ref{fig1}\textbf{f}). However, this distortion is independent of the position in the sample, such that we tentatively attribute it to residual strain in the thin films. Imperfections of the optical elements are less likely to cause this deviation, because we reproduce similar SHG patterns but with improved symmetry on a reference \(c\)-cut single crystal sample (Supplemental Figure S3).

Next, we utilize the SHG to spatially map the temperature evolution of the magnetic order. Figures \ref{fig2}\textbf{a-c} depict spatial maps of the SHG intensity at $\qty{160}{K}$, well below the Morin transition, recorded in co-polarization and at fixed excitation polarizations of \qty{30}{\degree}, \qty{150}{\degree} and \qty{270}{\degree}, corresponding to the high intensity lobes of the rotational anisotropy pattern. To highlight and compare spatial variations, each dataset is normalized individually. The sample appears divided into two distinct types of SHG domains characterized by high and low intensity, with a lateral size on the order of few micrometer. We tentatively assign these two domains to regions where the Néel vector is oriented parallel and antiparallel to the \(c\)-axis. In this scenario, the SHG intensity contrast arises from an interference between the electric quadrupole (even under time reversal) and the magnetic dipole (odd under time reversal) contribution to the nonlinear susceptibility (see Supplemental Material). Importantly, maps recorded at different orientations of the polarization appear almost identical except for an intensity normalization factor. Hence, we can conclude that the rotational SHG pattern is indeed spatially uniform as would be expected for the out-of-plane spin orientation below \TM and the corresponding trigonal symmetry. 

Above \TM, the Néel vector lies within the basal plane along specific in-plane directions. As a result, these SHG rotational anisotropy patterns deviate from the spatially uniform symmetry observed in the low-temperature phase (c.f. Fig. \ref{fig1}\textbf{f}). In particular, different magnetic domains related by $\qty{120}{\degree}$ rotations within the basal plane will give rise to distinct angular patterns. This becomes evident for the SHG maps recorded at \qty{270}{K} for \qty{30}{\degree}, \qty{150}{\degree}, and \qty{270}{\degree} (Fig. \ref{fig2}\textbf{d-f}). Even after intensity normalization, the three spatial maps remain qualitatively different. In the simplest case, there are now $2 \cdot 3 = 6$ distinct domain orientations of the N\'eel vector (either parallel or antiparallel to the three hexagonal lattice vectors indicated in Fig.~\ref{fig1}\textbf{b}). 

An important consequence of the domain sensitivity is that SHG can image the Morin transition. Figure \ref{fig3} shows spatial maps demonstrating an overall increase of the SHG intensity below \TM{}. While early works, using circularly polarized SHG, could not resolve any domain contrast in hematite, unlike for example \ce{Cr2O3} \cite{pisarev_nonlinear_1997}, our measurements at fixed linear polarization clearly resolve different spin re-orientations and domains. For example, for the position marked by the red cross in Fig.~\ref{fig3}\textbf{a-c}, SHG intensity is maximized in the easy-axis phase for $T < T_\mathrm{M}$, while it is minimized in the easy-plane phase for $T > T_\mathrm{M}$, i.e. the probed SHG domain is rotated away from the direction defined by the polarization. Consequently, we get a sharp decrease of the SHG signal (red data points in Fig. \ref{fig3}\textbf{d}) very close to the Morin temperature \TM{} determined by SQUID magnetometry (gray dashed line). Small deviations may be caused by local heating within the focused laser spot. Conversely, for the position marked by the orange cross, we start with minimum SHG in the easy-axis phase, and we end with maximum SHG in the easy-plane phase, i.e. the SHG domain is rotated exactly along the direction defined by the polarization. Consequently, we get a sharp increase of the SHG signal (orange data points in Fig. \ref{fig3}\textbf{d}) when crossing \TM. The locations defined by the brown and green crosses highlight exemplarily intermediate cases, where the transition is hence less pronounced.

While the identification of the Morin transition in SHG is fully consistent with a magnetic dipole contribution, it is not a definitive prove of an additional non-magnetic contribution, such as the assumed electric quadrupole term, which is crucial for the domain contrast mechanism. To this end, we also conducted SHG measurements as function of both in-plane and out-of-plane magnetic field. At room temperature, in the easy-plane phase, we measured the dependence of the SHG patterns on a fixed external in-plane field ($B_\mathrm{ip} = \qty{550 \pm 50}{mT}$). This field is expected to be sufficient to continuously rotate the N\'eel vector at room temperature, due to the rather weak in-plane anisotropy field \cite{fabian2011experimental, lebrun2019anisotropies}. Hereby, the N\'eel vector orients perpendicularly to the direction of the external magnetic field. The rotational patterns (cross-polarized) in Figs. \ref{fig4}\textbf{a-c} clearly suggest that the external magnetic field direction modifies the magnetic symmetry. Figure \ref{fig4}\textbf{d} depicts how the SHG intensity measured at a fixed sample position along the three principal lobes (green, red, and purple curves) changes indeed continuously as a function of the angle of $B_\mathrm{ip}$. Most importantly, the SHG pattern is \qty{360}{\degree}-symmetric in the external field (Fig. \ref{fig4}\textbf{d}). We interpret this finding such that the change in SHG contrast upon \qty{180}{\degree} rotation of the external field arises from a reversal of the N\'eel vector. We additionally verify the continuous magnetic field dependence (Supplemental Figure S5) on a second sample, where we also verify that the change persists even when the field is removed (Supplemental Figure S6) ruling out magnetic field induced second harmonic generation as the sole underlying mechanism \cite{fiebig_second_2001}. 

Finally, we demonstrate the magnetic origin of the SHG contrast also in the easy-axis phase for $T < T_\mathrm{M}$ using an out-of-plane field $B_\mathrm{oop}$ aligned along the $c$-axis (Fig. \ref{fig5}\textbf{a-c}). A large out-of-plane field induces a spin-flop transition of the magnetic moments aligned along the $c$-axis into the $c$-plane at approximately \qtyrange{6}{8}{T}\,\cite{lebrun2020long, dannegger_magnetic_2023}. Initially, at $B_\mathrm{oop} = \qty{0}{T}$ and $T = \qty{4.7}{K}$ (Fig. \ref{fig5}\textbf{a}), we observe a distinct separation into domains with high and low SHG intensity, originating from the two possible orientations of the N\'eel vector as discussed above (c.f. Fig. \ref{fig2}\textbf{a-c}). Upon increasing the field to $B_\mathrm{oop} = \qty{9}{T}$, the domain pattern changes, which we attribute as evidence for the spin-flop transition. Consistently, the SHG patterns recorded at different sample positions show distinct rotations (Figs. \ref{fig5}\textbf{d-f}). We note that we found small drifts during the measurement caused by slow heating of the sample stage under \qty{70.3}{mW} of infrared laser excitation. Therefore, we show data (triangles) that is symmetrized from the full \qty{360}{\degree} raw dataset. The resulting rotational patterns are well-described by a fit (gray line) using our symmetry-adapted source terms. Most importantly, as the magnetic field is cycled back to $B_\mathrm{oop} = \qty{0}{T}$ (Fig. \ref{fig5}\textbf{c}), the final spatial pattern remains changed as compared to the start of the measurement (Fig. \ref{fig5}\textbf{a}). This remanent change in image contrast upon cycling of the out-of-plane field across the spin-flop transition proves the N\'eel vector as the origin of the SHG contrast also in the easy-axis phase. Furthermore, it demonstrates that the magnetic domain configuration in the easy-axis phase can change due to magnetic field induced spin-reorientation transitions. 

\section{Conclusion}

Overall, our findings demonstrate that standard SHG microscopy can be used to image magnetic domains in (0001)-oriented altermagnetic hematite films. In the easy-plane phase (above \TM), an obvious contrast mechanism for linearly polarized excitation arises from the in-plane orientations of the magnetic dipole contribution, which is tied to the N\'eel vector orientation along three equivalent lattice vectors. This results in two-fold symmetric SHG polarization patterns, which are well-described by magnetic dipole and electric quadrupole source terms as derived from a symmetry analysis of the tensor elements. 
We note that, in principle, a surface or interface electric dipole term could explain our thin film findings as well. However, control experiments on a bulk reference crystal show comparable SHG contrast (Supplemental Figure S3), such that we base our symmetry analysis on the lowest order bulk terms. Additionally, we reproduce the domain structure and the in-plane magnetic field dependence at room temperature on a second thin film sample (Supplemental Figures S4, S5, and S6). Lastly, control experiments with a low numerical aperture objective lens rule out that the magnetic contrast is caused by the interplay of out-of-plane electric and in-plane electric fields within the focal spot (Supplemental Figure S7).

In the easy-axis phase (below \TM), we can image the orientation of the N\'eel vector in the thin films by a simple binary intensity contrast, where we find domains with typical lateral sizes on the order of several micrometer. In principle, this would also allow us to fully image the magnetic order in the easy-plane phase, similar to recent works employing x-ray magnetic contrast imaging \cite{galindez2025revealing}. However, the more complicated domain structure with six, rather than two, possible orientations of the N\'eel vector impedes simple intensity thresholding. Rather a full rotational SHG pattern would need to be resolved at each position. Furthermore, potential contributions from sub-diffraction scale magnetic features, such as smaller domains, domain walls, or merons \cite{galindez2025revealing, wornle2021}, and small residual strains may distort the SHG patterns requiring careful quantitative calibration, which is beyond the scope of the current work.

Nevertheless, we already gain important insights into the magnetic structure without resolving the full domain picture. From the out-of-plane magnetic field measurements (c.f. Fig. \ref{fig5}), we can conclude that domains are not pinned after a cycled spin-flop transition measured at $T = \qty{4.7}{K}$. This stands in contrast to temperature dependent measurements. There, we find that the domain pattern at $T = \qty{160}{K}$ is largely pinned, with the largest observed domains being on the order of \qtyrange{5}{10}{\micro\meter} (c.f. Fig. \ref{fig2} and Fig. \ref{fig5}), whereas the domain pattern at $T = \qty{270}{K}$ changes continuously upon repeated cycling between the respective temperatures (Supplemental Figure S7). In light of the renewed interest in hematite as an altermagnetic candidate material, such magnetic imaging can be very relevant for interpreting transport data on mesoscopic samples that may average across domains and may be sensitive to absence or presence of domain reorientation and pinning across multiple measurement cycles. Finally, higher harmonic generation, or nonlinear currents in general \cite{kiemle_light-field_2020, dong2025, sivianes2025}, can be a symmetry-selective tool to directly probe spin-splittings or hidden symmetries tied to altermagnetic order \cite{zhao_evidence_2016, ma_probing_2025, wu2025optical, usachev2026nonlinear}.

\section{Methods}\label{sec11}

\textbf{Raman and SHG microscopy.} Second harmonic generation and Raman microscopy were implemented with a \qty{1064}{nm} picosecond pulsed excitation laser ($P_\mathrm{laser} \approx$ \qtyrange{10}{200}{mW}) and a \qty{532}{nm} cw excitation laser ($P_\mathrm{laser} \approx$ \qtyrange{1}{5}{mW})), respectively. The lasers were coupled into a commercial confocal microscope (WITec Alpha 300 R). For measurements under ambient conditions and for temperature dependent measurements, we used 100x (0.9 NA) and 63x (0.7 NA, with cover glass correction) objectives, respectively. Temperature dependent measurements were conducted in a flow cryostat under vacuum conditions (base pressure $<\qty{1e-6}{mbar}$). The linear polarization of the excitation light was controlled by an achromatic half-wave plate. For all measurements the excitation spot was diffraction limited. The reflected light passed the same half-wave plate, and it was collected confocally by a dispersive grating spectrometer (Raman spectra \qty{1800}{groves/mm} grating, SHG \qty{300}{groves/mm} grating) and a front-illuminated CCD. For low-temperature SHG in an out-of-plane magnetic field, we used a confocal cryoRaman platform (WITec microscope and Attocubbe cryostat) with 69.2x low-temperature objective (0.81 NA), 1064 nm excitation ($P_\mathrm{laser} \approx$ \qty{70}{mW}), \qty{600}{groves/mm} grating, and a back illuminated CCD. For all SHG measurements, the detection was co-polarized or cross-polarized as noted in the corresponding figure description.\\
\textbf{Thin film growth.} Epitaxial \hema{} films were grown via pulsed laser deposition on (0001)-oriented, single crystalline sapphire (\ce{Al2O3}) substrates. The growth process was carried out in an oxygen atmosphere with a partial pressure of \qty{2.5e-5}{bar}, while the substrate temperature was kept at \qty{320}{\degree C}. The laser fluence at the polycrystalline \hema{} target was \qty{2.5}{Jcm^{-2}}, and the pulse repetition rate was set to \qty{2}{Hz}. The magnetization of the films was determined by SQUID magnetometry. The in-plane crystallographic direction of the films with respect to the growth substrate was determined from XRD analysis. Further details on the growth can be found in Ref. \cite{scheufele_impact_2023}.

\section*{Competing interests}
The authors declare no competing interests.

\section*{Acknowledgements}
Work was supported by the Deutsche Forschungsgemeinschaft (DFG, German Research Foundation) via the Munich Center for Quantum Science and Technology No. (MCQST)-EXC-2111-390814868. A.W.H. acknowledges the excellence cluster e-conversion No. EXC-2089/1-390776260. A.W.H and H.H. acknowledge the Munich Quantum Valley K1 and K9, which the Bavarian state government supports with funds from the Hightech Agenda Bayern Plus as well as the One Munich Strategy Forum—EQAP. K.M. and M.A. acknowledge funding by the European Union's Horizon Europe research and innovation program under Grant No. 101171325 (POSA). H.H. acknowledges the Transregio ConQuMat (TRR 360 – 492547816). M.A., S.G., and H.H. acknowledge funding via the DFG priority programme SPP 2558.  P.S. thanks dtec.bw—Digitalization and Technology Research Center of the Bundeswehr for support (project VITAL-SENSE). dtec.bw was funded via the German Recovery and Resilience Plan by the European Union (NextGenerationEU). We thank Giancarlo Soavi for his advice regarding the magnetic point groups of hematite.

\section*{Author contributions}
C.K., A.H., H.H., M.A., and S.G. conceptualized the study. K.M. and S.G. grew the samples and measured crystallographic orientation and magnetization. H.M. and J.S. carried out the optical measurements, analyzed the data, and carried out the theoretical symmetry analysis. J.P. and P.S. provided low-temperature second harmonic generation measurements with magnetic field. H.M., J.S., and C.K. wrote the manuscript with input from all authors. H.M. and J.S. contributed equally. All authors reviewed the manuscript.

\printbibliography



\section*{Supplementary material}

## Temperature dependent Raman spectra

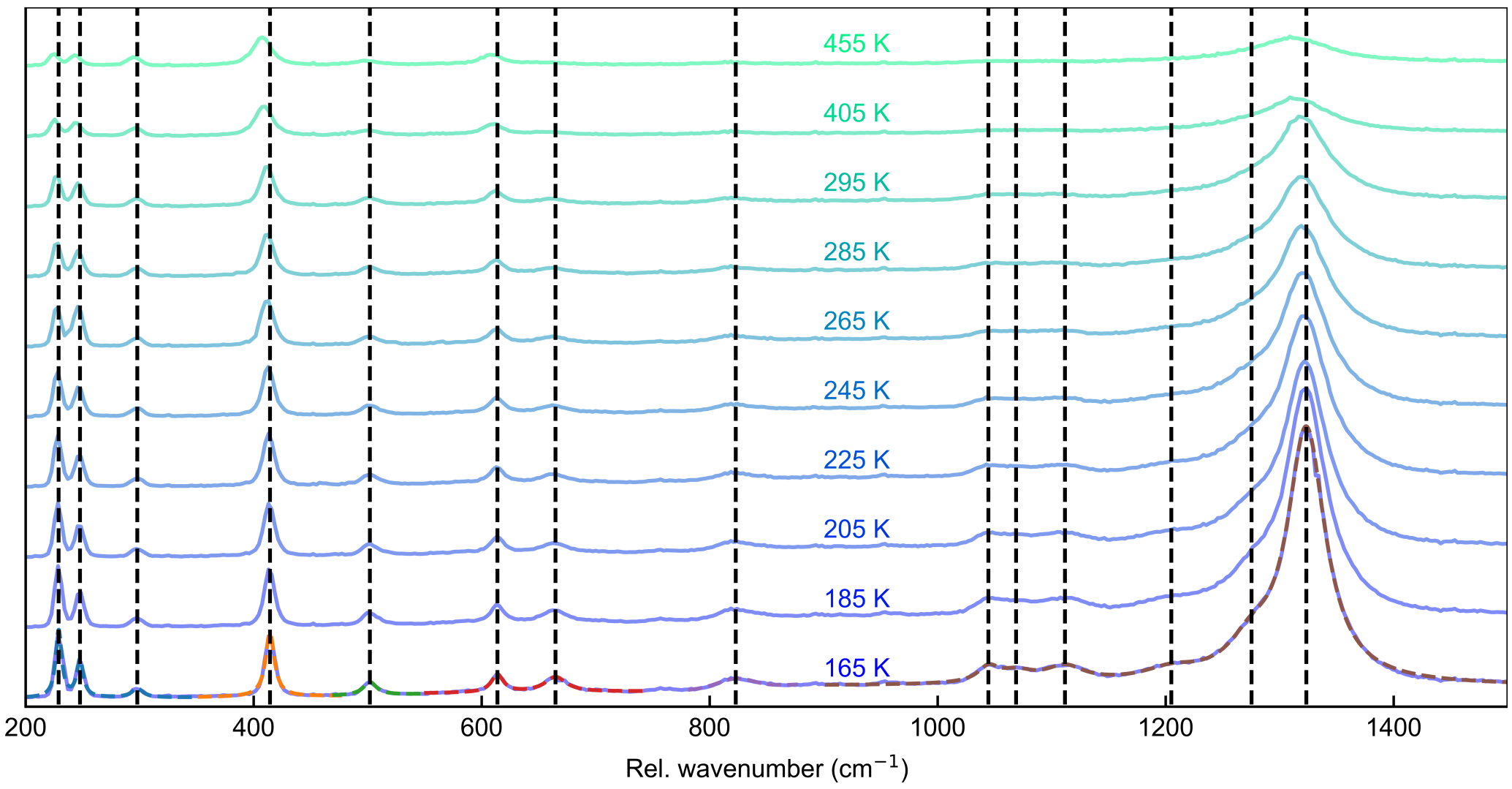


| | $A_{1g(1)}$ | $E_{g(1)}$ | $E_{g(2)} + E_{g(3)}$ | $A_{1g(2)}$ | $E_{g(4)}$ | $E_{g(5)}$ | $E_u$ | *Magnon* | $E_{u,ot}$ |
|---|---|---|---|---|---|---|---|---|---|
| **165 K** | 229.00 | 247.72 | 298.04 | 414.30 | 502.01 | 613.82 | 664.83 | 822.96 | 1323.42 |
| **185 K** | 228.69 | 247.46 | 297.72 | 413.90 | 502.00 | 613.62 | 663.97 | 822.66 | 1323.03 |
| **205 K** | 228.35 | 247.28 | 297.88 | 413.65 | 502.07 | 613.55 | 663.64 | 821.85 | 1322.57 |
| **225 K** | 228.05 | 247.16 | 298.16 | 413.20 | 502.01 | 613.23 | 663.10 | 821.65 | 1322.09 |
| **245 K** | 227.75 | 246.81 | 298.02 | 412.74 | 502.03 | 613.05 | 662.87 | 821.04 | 1321.05 |
| **265 K** | 227.29 | 245.95 | 297.90 | 411.81 | 500.68 | 612.97 | 662.78 | 820.12 | 1320.53 |
| **285 K** | 227.22 | 246.28 | 297.64 | 411.94 | 501.76 | 612.38 | 662.04 | 820.64 | 1319.82 |
| **295 K** | 227.14 | 246.25 | 297.48 | 411.75 | 501.45 | 612.13 | 661.13 | 819.40 | 1319.47 |
| **405 K** | 225.39 | 244.23 | 296.37 | 408.90 | 500.10 | 609.94 | 658.33 | 819.17 | 1314.61 |
| **455 K** | 224.51 | 243.42 | 294.97 | 407.49 | 498.52 | 608.48 | 655.39 | 818.82 | 1312.25 |

Figure 1: **Raman shifts of the thin film sample for temperatures from** 165 K **to** 455 K**.** The table shows the fitted peak positions of the Raman modes according to Ref. [1]. The spectra were detected without an analyzer (unpolarized).

## Group theory SHG calculations in hematite

In reflection geometry for c-plane oriented films, the incident electric field propagates along the c-axis and we assume a transverse wave: $\vec{E} = (E_x, E_y, 0)^T = E_0(\cos\phi, \sin\phi, 0)^T$. We model the SHG, which is detected in reflection geometry, using the following contributions from bulk hematite to the effective nonlinear polarization $\mathbf{P}^{(2)}_{\text{eff}}(2\omega)$:

$$\mathbf{P}^{(2)}_{\text{eff}}(2\omega) = \mathbf{P}^{(2)}(2\omega) + \left(\frac{c}{2i\omega}\right)\nabla\times\mathbf{M}^{(2)}(2\omega) - \nabla\mathbf{Q}^{(2)}(2\omega). \tag{1}$$

where $\mathbf{P}^{(2)}$, $\mathbf{M}^{(2)}$, and $\mathbf{Q}^{(2)}$ denote the electric dipole (ED), magnetic dipole (MD), and electric quadrupole (EQ), respectively. Hematite $\alpha$-$Fe_2O_3$ possesses a corundum type crystal structure (rhombohedral space group $D^6_{3d}$). Inversion symmetry in hematite is not broken by the AFM ordering of spins below $T_{\text{N}}$= 950 K, hence the electric dipole polarization term vanishes. The remaining terms in Eq.1 are quadratic functions of the fundamental light field, and one can write

$$M_i(2\omega) = (\epsilon_0 c/n)\chi^m_{ijk}E_j(\omega)E_k(\omega) \tag{2}$$
$$Q_{ij}(2\omega) = \epsilon_0\chi^q_{ijkl}E_k(\omega)E_l(\omega). \tag{3}$$

Below the Morin transition, the magnetic point group is $\bar{3}'m'$ [2]. Above $T_{\text{M}}$ spins lie in the easy plane and the corresponding magnetic point group is $2'/m'$[3, 4]. The non-magnetic space group is $\bar{3}m$, in both cases. To calculate the SHG response, the effective polarization source term is derived, and the co-polarized ($I_{XX}$) and cross-polarized ($I_{YX}$) detection channels are calculated as follows:

$$\begin{aligned}
S_i &= \mu_0\frac{\partial^2 P_i}{\partial t^2} + \mu_0\,\varepsilon_{ijk}\,\nabla_j\frac{\partial M_k}{\partial t} - \mu_0\,\nabla_j\frac{\partial^2 Q_{ij}}{\partial t^2},\\
\hat{e}_{XX} &= (\cos\phi, \sin\phi, 0),\\
\hat{e}_{YX} &= (-\sin\phi, \cos\phi, 0),\\
A_{XX} &= \hat{e}_{XX}\cdot\mathbf{S} = \cos\phi S_x + \sin\phi S_y,\\
A_{YX} &= \hat{e}_{YX}\cdot\mathbf{S} = -\sin\phi S_x + \cos\phi S_y,\\
I_{XX}(\phi) &\propto |A_{XX}(\phi)|^2,\\
I_{YX}(\phi) &\propto |A_{YX}(\phi)|^2.
\end{aligned} \tag{4}$$

Below $T_{\text{M}}$, the non-zero elements of the magnetization dependent term $\mathbf{M}^{(2)}$ are given by [5]:

$$\begin{aligned}
\chi^m_{xxx} &= xxx,\\
\chi^m_{xyy} &= \chi^m_{yxy} = \chi^m_{yyx} = -xxx.
\end{aligned} \tag{5}$$

The polarization components are:

$$\begin{aligned}
M_x &= \left(\frac{\varepsilon_0 c}{n}\right)\chi^m_{xxx}\,E_xE_x - \left(\frac{\varepsilon_0 c}{n}\right)\chi^m_{xxx}\,E_yE_y,\\
M_y &= -2\left(\frac{\varepsilon_0 c}{n}\right)\chi^m_{xxx}\,E_xE_y,\\
M_z &= 0.
\end{aligned} \tag{6}$$

Following Eq.4 we get:

$$\begin{aligned}
I_{XX} &\propto |\chi^m_{xxx}|^2\sin^2(3\phi),\\
I_{YX} &\propto |\chi^m_{xxx}|^2\cos^2(3\phi).
\end{aligned} \tag{7}$$

Analogously, we derive the magnetic terms above the Morin transition, where the symmetry is reduced to $2'/m'$:

$$\begin{aligned}
\chi^m_{xxx} &= xxx,\\
\chi^m_{yyx} &= \chi^m_{yxy} = \chi^m_{xyy} = yyx,\\
\chi^m_{zzx} &= \chi^m_{zxz} = \chi^m_{xzz} = zzx,\\
\chi^m_{xyz} &= \chi^m_{yxz} = \chi^m_{xzy} = \chi^m_{yzx} = \chi^m_{zxy} = \chi^m_{zyx} = xyz.
\end{aligned} \tag{8}$$

The polarization components are:

$$
\begin{aligned}
M_x &= \left(\frac{\varepsilon_0 c}{n}\right) \chi^m_{xxx} E_x E_x + \left(\frac{\varepsilon_0 c}{n}\right) \chi^m_{xyy} E_y E_y, \\
M_y &= \left(\frac{2\varepsilon_0 c}{n}\right) \chi^m_{xyy} E_x E_y, \\
M_z &= \left(\frac{2\varepsilon_0 c}{n}\right) \chi^m_{xyz} E_x E_y.
\end{aligned} \tag{9}
$$

Following Eq.4 we get:

$$
\begin{aligned}
I_{XX} &\propto |-2\chi^m_{xyy}\cos^2\phi\sin\phi + \chi^m_{xxx}\sin\phi\cos^2\phi + \chi^m_{xyy}\sin^3\phi|^2, \\
I_{YX} &\propto |3\chi^m_{xyy}\sin^2\phi\cos\phi + \chi^m_{xxx}\cos^3\phi|^2.
\end{aligned} \tag{10}
$$

The second-order polarization acquires an electric-quadrupole (EQ) contribution driven by the gradient of the optical field:

$$
P^q_i(2\omega) = \chi^q_{ijkl}\, E_j(\omega)\, \partial_k E_l(\omega). \tag{11}
$$

The EQ susceptibility $\chi^q_{ijkl}$ is therefore a 4[th]-rank polar tensor. Neumann's principle requires invariance under every symmetry operation $R \in G$:

$$
\chi^q_{ijkl} = R_{ii'} R_{jj'} R_{kk'} R_{ll'}\, \chi^q_{i'j'k'l'} \quad \text{with} \quad i,j,k,l \in \{x,y,z\}, \qquad \forall R \in G. \tag{12}
$$

The generators of $\bar{3}m$ ($D_{3d}$) with the trigonal axis along $z$ and a mirror operation $\sigma_v \perp x$ are:

$$
C_3 = \begin{pmatrix} -\frac{1}{2} & -\frac{\sqrt{3}}{2} & 0 \\ \frac{\sqrt{3}}{2} & -\frac{1}{2} & 0 \\ 0 & 0 & 1 \end{pmatrix}, \quad \sigma_v = \begin{pmatrix} -1 & 0 & 0 \\ 0 & 1 & 0 \\ 0 & 0 & 1 \end{pmatrix}, \quad \bar{1} = -\mathbb{1}. \tag{13}
$$

The independent components of $\chi^q_{ijkl}$ are fond by solving this system of constitutive relations

$$
\chi^q_{ijkl} - R_{ii'} R_{jj'} R_{kk'} R_{ll'}\, \chi^q_{i'j'k'l'} = 0. \tag{14}
$$

This can be rewritten as:

$$
\mathbf{A}\mathbf{x} = 0 \quad \text{with} \quad \mathbf{x} = \begin{bmatrix} \chi^q_{xxxx} \\ \chi^q_{xxxy} \\ \vdots \\ \chi^q_{zzzz} \end{bmatrix} \tag{15}
$$

It is therefore sufficient to find the vectors $\mathbf{v}^{(n)}$ spanning the nullspace of $\mathbf{A}$. The components of $\chi^q$ are then obtained by multiplying the free parameters $a_n$ with the vectors $\mathbf{v}^{(n)}$[5, 6]:

$$
\sum_n a_n \mathbf{v}^{(n)} = \mathbf{x} \tag{16}
$$

This reduces the tensor to these components with 14 independent variables:

$$
\begin{aligned}
&\chi^q_{xxxx} = xxxx, && \chi^q_{xxyy} = xxyy, && \chi^q_{xxyz} = -xxyz, && \chi^q_{xxzy} = -xxzy, \\
&\chi^q_{xxzz} = xxzz, && \chi^q_{xyxy} = xyxy, && \chi^q_{xyxz} = -xxyz, && \chi^q_{xyyx} = -xyxy - xxyy + xxxx, \\
&\chi^q_{xyzx} = -xxzy, && \chi^q_{xzxy} = -xzxy, && \chi^q_{xzxz} = xzxz, && \chi^q_{xzyx} = -xzxy, \\
&\chi^q_{xzzx} = xzzx, && \chi^q_{yxxy} = -xyxy - xxyy + xxxx, && \chi^q_{yxxz} = -xxyz, && \chi^q_{yxyx} = xyxy, \\
&\chi^q_{yxzx} = -xxzy, && \chi^q_{yyxx} = xxyy, && \chi^q_{yyyy} = xxxx, && \chi^q_{yyyz} = xxyz, \\
&\chi^q_{yyzy} = xxzy, && \chi^q_{yyzz} = xxzz, && \chi^q_{yzxx} = -xzxy, && \chi^q_{yzyy} = xzxy, \\
&\chi^q_{yzyz} = xzxz, && \chi^q_{yzzy} = xzzx, && \chi^q_{zxxy} = -zxxy, && \chi^q_{zxxz} = zxxz, \\
&\chi^q_{zxyx} = -zxxy, && \chi^q_{zxzx} = zxzx, && \chi^q_{zyxx} = -zxxy, && \chi^q_{zyyy} = zxxy, \\
&\chi^q_{zyyz} = zxxz, && \chi^q_{zyzy} = zxzx, && \chi^q_{zzxx} = zzxx, && \chi^q_{zzyy} = zzxx, \\
&\chi^q_{zzzz} = zzzz
\end{aligned} \tag{17}
$$

For linearly polarized light incident parallel to the z axis this can be simplified further since $E_z = \partial_x E_x = \partial_x E_y = \partial_y E_x = \partial_y E_y = 0$. The remaining result is therefore:

$$\begin{aligned} P_x &= -xxzy\,\epsilon_0\,(E_x\partial_z E_y + E_y\partial_z E_x), \\ P_y &= -xxzy\,\epsilon_0\,(E_x\partial_z E_x - E_y\partial_z E_y), \\ P_z &= zxzx\,\epsilon_0\,(E_x\partial_z E_x + E_y\partial_z E_y) \end{aligned} \tag{18}$$

With the prior definition of the the incident light electrical field the resulting in-plane moment is projected on the axis parallel and perpendicular to the incident polarization following 4.

$$\begin{aligned} I_{xx} &\propto |xxzy\sin(3\phi)|^2, \\ I_{xy} &\propto |xxzy\cos(3\phi)|^2 \end{aligned} \tag{19}$$

The resulting EQ terms match the findings in Ref.[7].

The sum of the MD and EQ fields yields the angular dependence of the in-plane SHG field, which in turn is used to derive the angular dependence of the SHG intensity, as it is measured in the experiment.

$$\begin{aligned} I_{XX}^{T<T_M} &\propto |\chi^m_{xxx}\sin 3\phi + \chi^q_{xxzy}\sin 3\phi|^2, \\ I_{YX}^{T<T_M} &\propto |\chi^m_{xxx}\cos 3\phi + \chi^q_{xxzy}\cos 3\phi|^2, \\ I_{XX}^{T>T_M} &\propto |-2\chi^m_{xyy}\cos^2\phi\sin\phi + \\ &\quad + \chi^m_{xxx}\sin\phi\cos^2\phi + \chi^m_{xyy}\sin^3\phi + \chi^q_{xxzy}\sin 3\phi|^2, \\ I_{YX}^{T>T_M} &\propto |3\chi^m_{xyy}\sin^2\phi\cos\phi + \chi^m_{xxx}\cos^3\phi + \chi^q_{xxzy}\cos 3\phi|^2. \end{aligned} \tag{20}$$

Supplemental Fig. S2 shows typical representations of the calculated co- and cross-detected SHG intensities $I_{XX}(\phi)$ and $I_{YX}(\phi)$ for a chosen set of parameters:

$$\begin{aligned} \frac{\chi^{2'/\mathrm{m}'}_{xxx}}{\chi^{\bar{3}'\mathrm{m}'}_{xxx}} &= \frac{2}{9}, \\ \frac{\chi^{2'/\mathrm{m}'}_{xyy}}{\chi^{2'/\mathrm{m}'}_{xxx}} &= -\frac{3}{2}, \\ \chi^q_{xxzy} &= \frac{1}{2}. \end{aligned} \tag{21}$$

Importantly, the interference between the MD and EQ term can explain the experimentally observed domains. In the low-temperature AFM phase, two different domains can be realized, which only distinguish by a spin inversion along the $C_3$-axis: $(-++-) \to (+--+)$. This inversion changes the sign of $\chi^{\mathrm{m}}_{xxx}$, but the coexisting EQ contributions remains unchanged. The resulting interference leads to a measurable SHG intensity contrast allowing for differentiation of the two antiferromagnetic domains (compare light and dark SHG pattern in SI Fig. 2d).

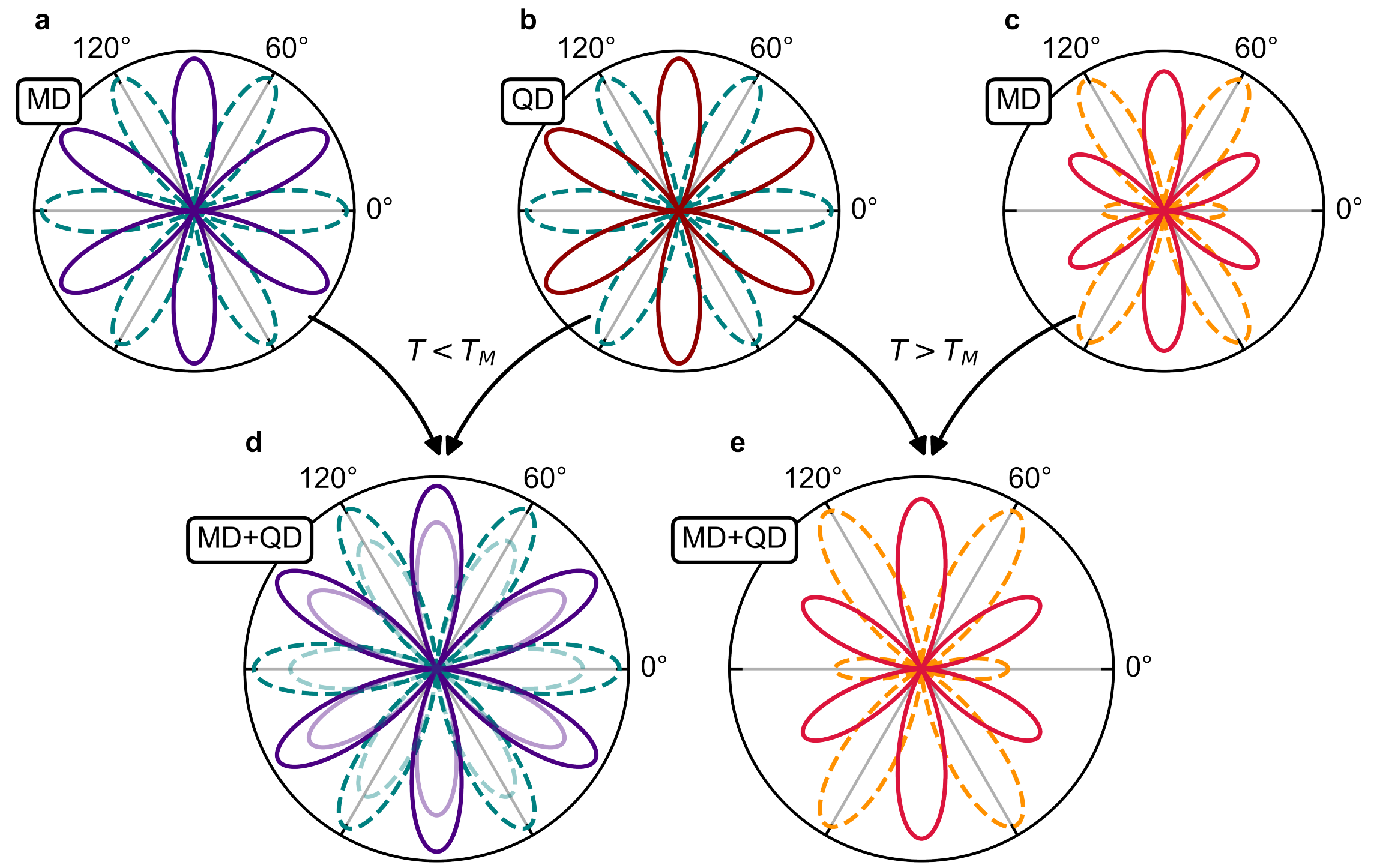


Figure 2: **Calculated SHG intensity.** **a** For the magnetic dipole below $T_{\mathrm{M}}$(co-polarized as solid lines, cross-polarized as dashed lines), **b** the electric quadrupole, and **c** the magnetic dipole above $T_{\mathrm{M}}$. The resulting patterns from the interference of these contributions are depicted in **d** below $T_{\mathrm{M}}$ and **e** above $T_{\mathrm{M}}$.

## Bulk single crystal measurements

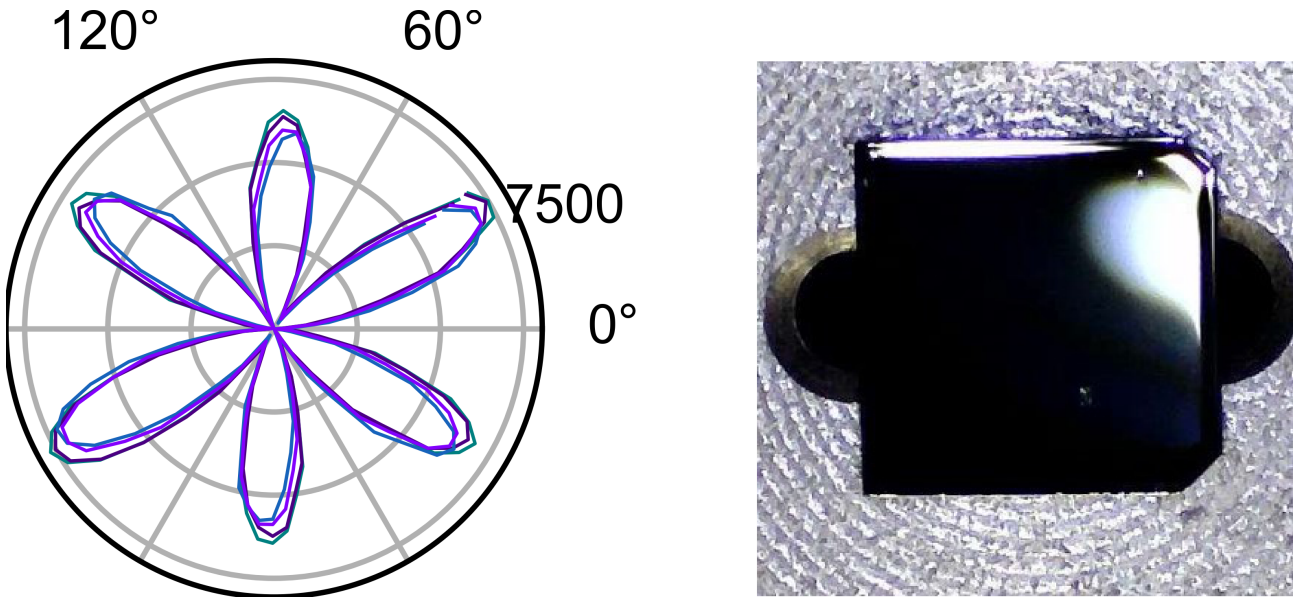


Figure 3: **SHG measured on a c-cut hematite single crystal**. The polarization-resolved SHG patterns were recorded in co-polarized detection below $T_M$. The different colors correspond to four different spatial positions on the single crystal.

## Measurements on a second thin film sample

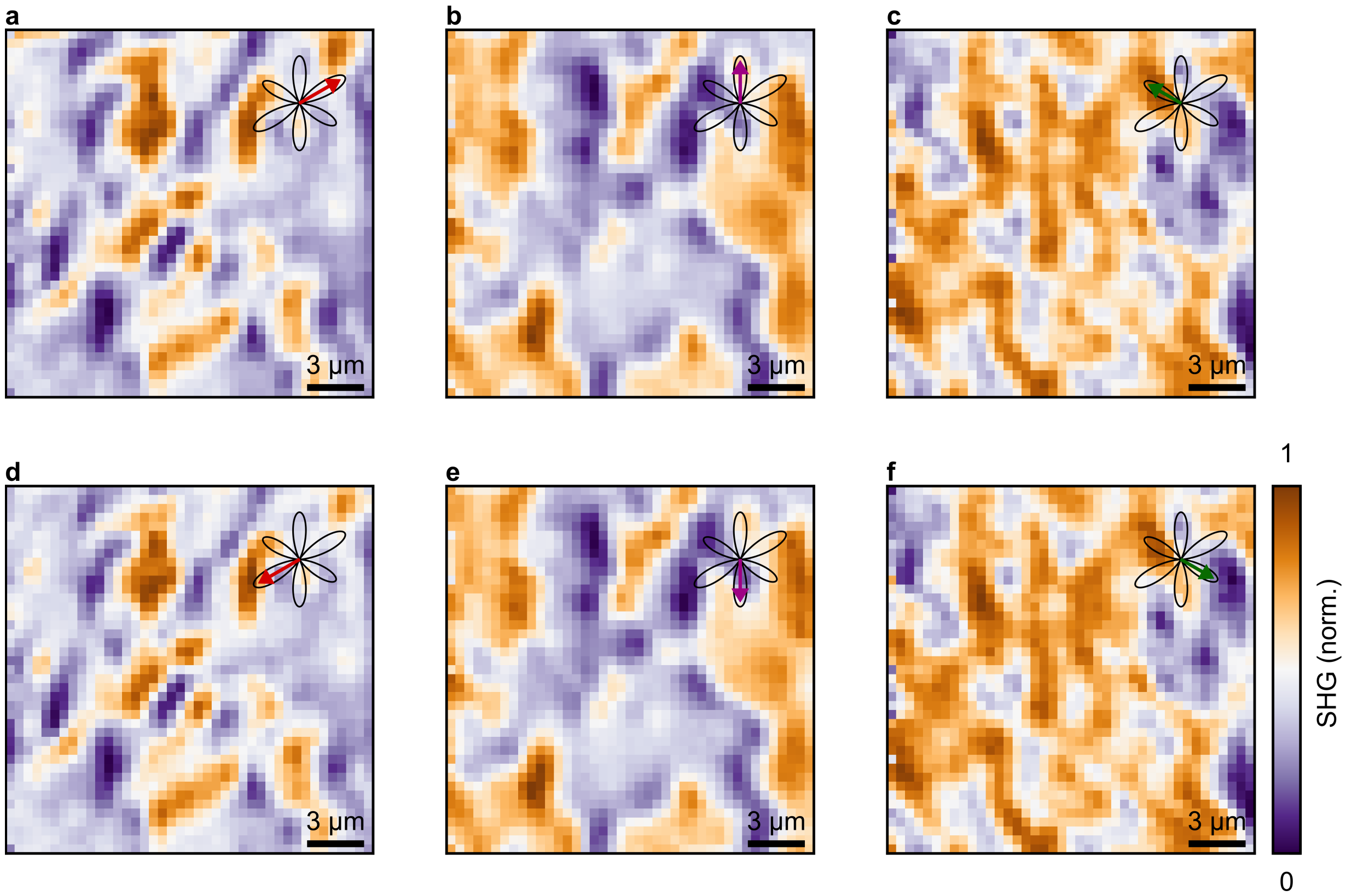


Figure 4: **Spatial mapping of SHG above the Morin transition on a second sample. a-c** show the spatial SHG intensities for analyzer positions chosen to coincide with the lobe positions. The directions are depicted in the top right corners. All measurements are in co-polarized detection. **d-f** present the maps with the same polarization as a-c but with 90°-rotated half-wave-plate.

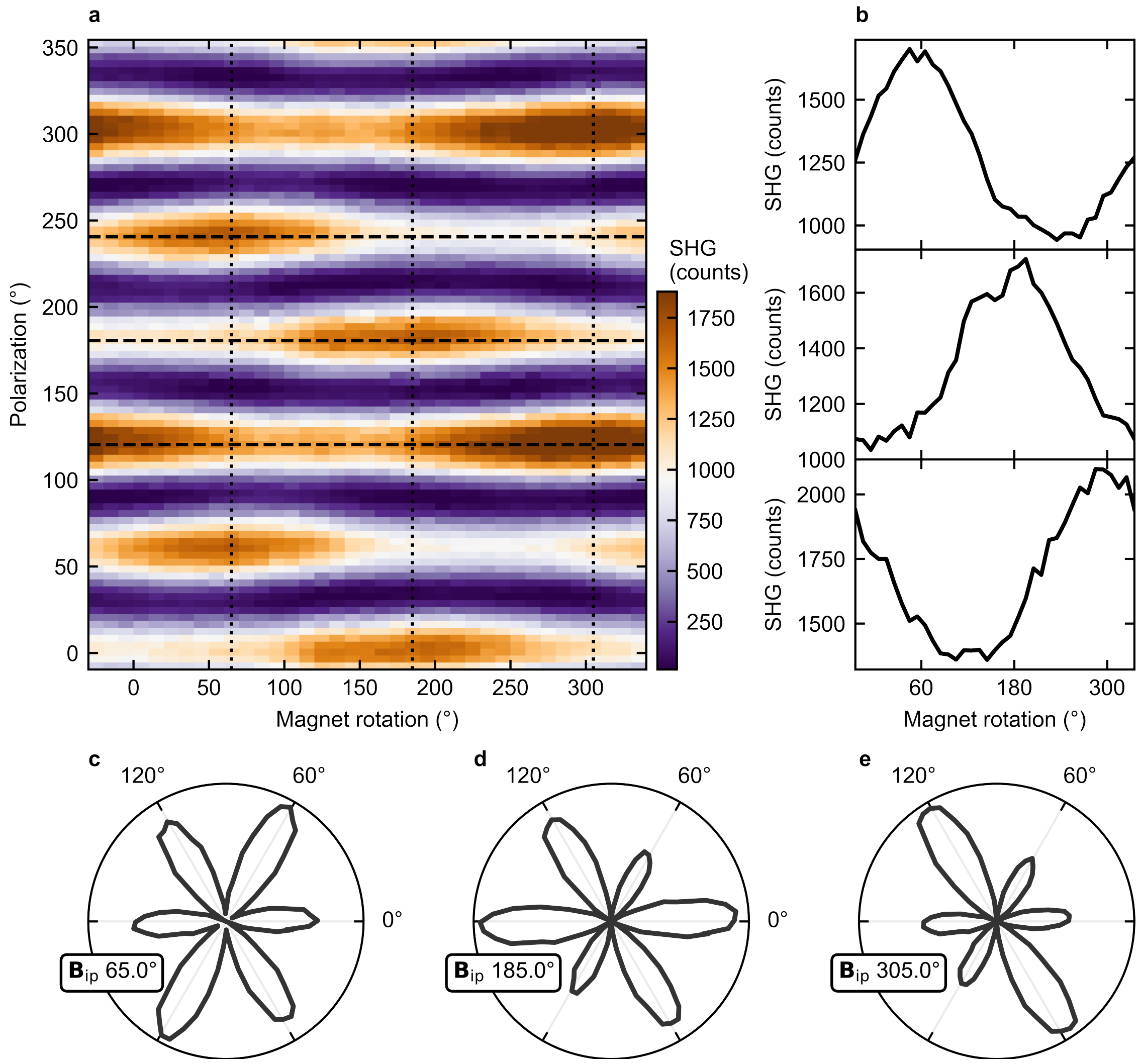


Figure 5: **Polarization- and in-plane magnetic field dependence of SHG intensity on the second sample.** The in-plane field $B_{\mathrm{ip}}$ is rotated from 0° to 360° using a rotatable array of permanent magnets. The field at the sample position is 500 mT to 600 mT. The entire parameter space is mapped in **a**, revealing 120°-shifted sinusoidal intensity variations in **b** for every lobe (dashed lines in a). **c-e** show cross-polarized SHG patterns for the three intensity maxima (dotted lines in a) .

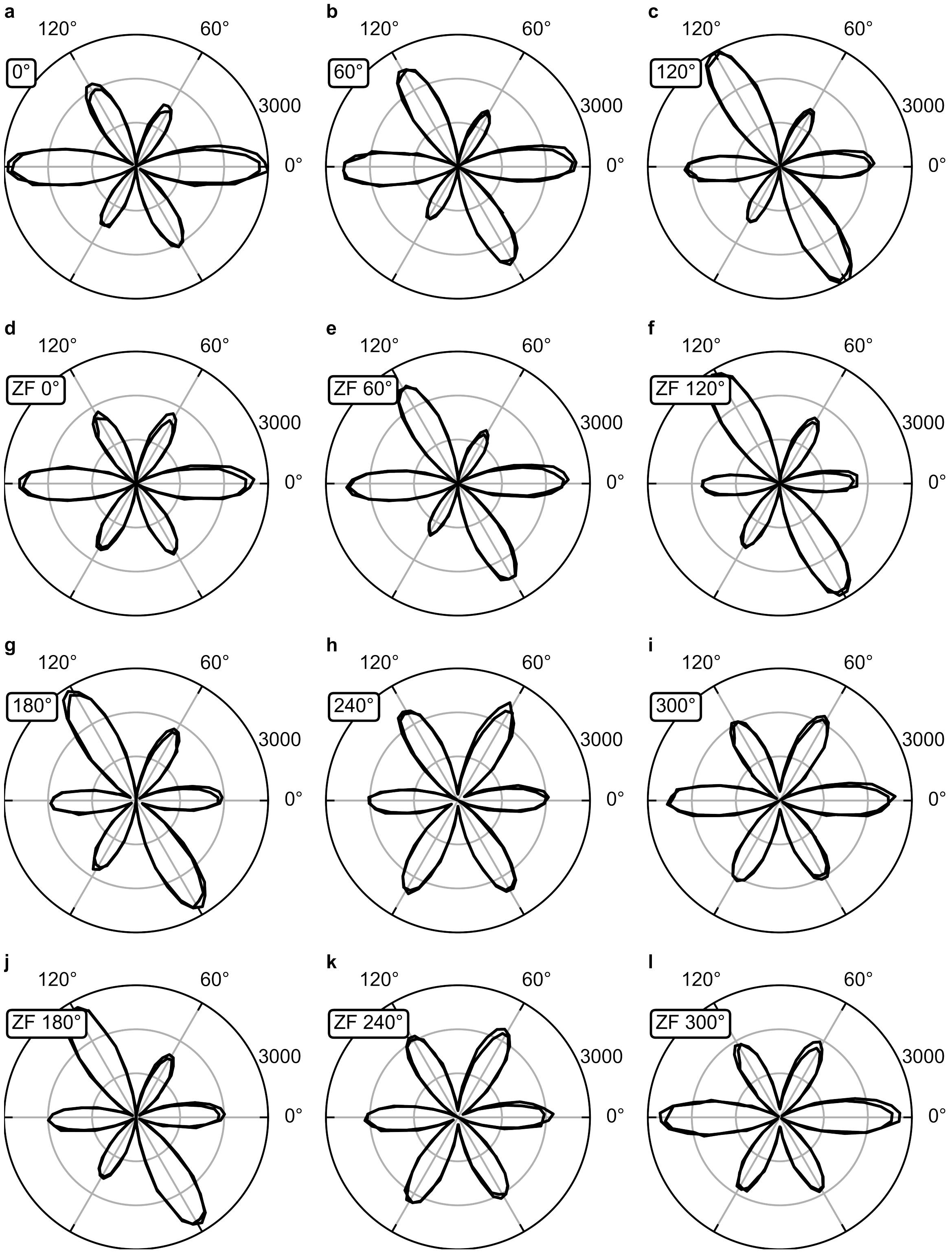


Figure 6: **Polarization-dependent SHG intensities at finite and zero in-plane magnetic field. a,b,c** show cross-polarized SHG on the second sample as the magnetic field is present (rotatable array of permanent magnets, 500 mT to 600 mT) . In **d,e,f** the in-plane magnetic field was first set to the corresponding in-plane angle and then set to zero by moving the sample out of the magnetic field plane (ZF 0 mT). The three magnetic field directions where the SHG is largest are shown. **g,h,i** visualize the polarization-dependence of the SHG signal when the field is further rotated and with respect to a,b,c reversed in direction (and at zero field **j,k,l**)

## SHG contrast using different numerical apertures

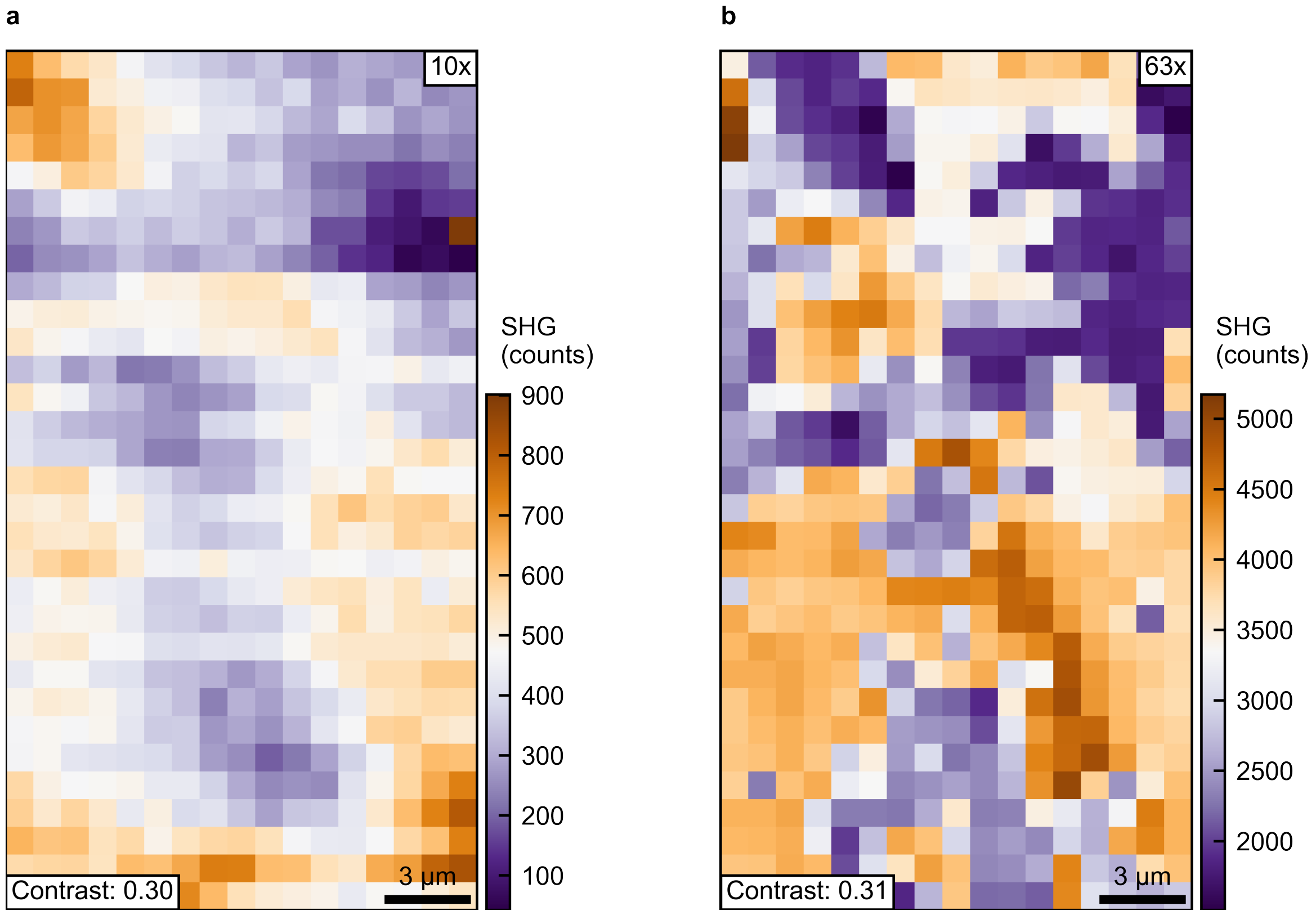


Figure 7: **Cross-polarized spatial SHG map at** 80 K **for NA=0.25 and NA=0.75 a** Map measured with a 10x objective (NA=0.25). **b** Map measured with a 63x objective (NA=0.75) at the same location as in a. Contrast is calculated by comparing the upper and lower 5% to 30% of the SHG intensities detected in the respective maps.

## Measurement of SHG upon temperature cycling

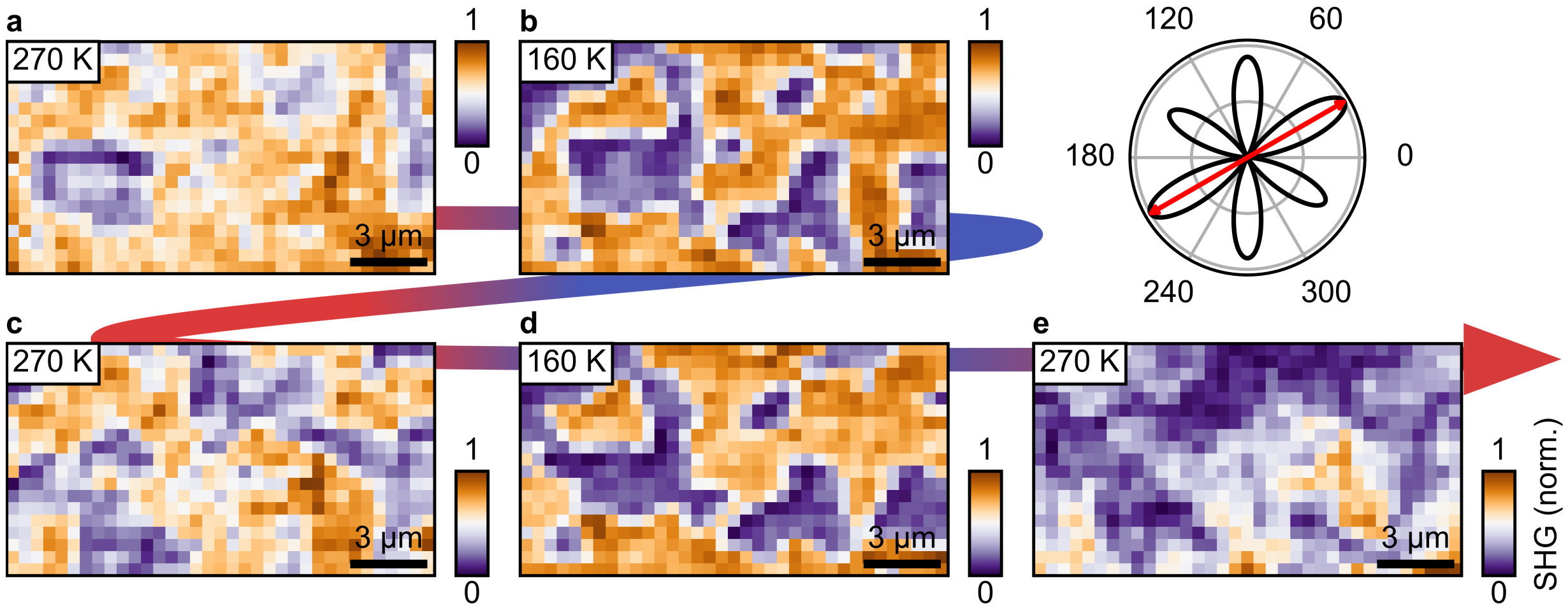


Figure 8: **Evolution of magnetic domains upon temperature cycling**. The images show spatial mapping of SHG intensity in co-polarized detection. **a** Initial room temperature measurement. **b** After subsequent cooling to 160 K. **c** After subsequent heating to 270 K. **d** After subsequent cooling to 270 K. **e** After final heating to 270 K. The measurement series reveals a strong reorganization of the room temperature domain structure, whereas the domain structure below $T_{\mathrm{M}}$ changes only slightly.



\end{document}